# Changes in the near edge X-ray absorption fine structure of hybrid organic-inorganic resists upon exposure

Roberto Fallica[*,a,†], Benjamin Watts[a], Benedikt Rösner[a], Gioia Della Giustina[b], Laura Brigo[b], Giovanna Brusatin[b], Yasin Ekinci[a]

[a] Paul Scherrer Institute, 5232 Villigen PSI, Switzerland
[b] Department of Industrial Engineering, University of Padua, via Marzolo 9, Padova, Italy
* Corresponding Author: roberto.fallica@psi.ch; phone 0041 56 310 4578; fax 0041 56 310 2111; www.psi.ch
† Currently at IMEC, Kapeldreef 75, 3001 Leuven, Belgium

**ABSTRACT**

We report on the near edge X-ray absorption fine structure (NEXAFS) spectroscopy of hybrid organic-inorganic resists. These materials are nonchemically amplified systems based on Si, Zr, and Ti oxides, synthesized from organically modified precursors and transition metal alkoxides by a sol-gel route and designed for ultraviolet, extreme ultraviolet and electron beam lithography. The experiments were conducted using a scanning transmission X-ray microscope (STXM) which combines high spatial-resolution microscopy and NEXAFS spectroscopy. The absorption spectra were collected in the proximity of the carbon edge (~ 290 eV) before and after *in situ* exposure, enabling the measurement of a significant photo-induced degradation of the organic group (phenyl or methyl methacrylate, respectively), the degree of which depends on the configuration of the ligand. Photo-induced degradation was more efficient in the resist synthesized with pendant phenyl substituents than it was in the case of systems based on bridging phenyl groups. The degradation of the methyl methacrylate group was relatively efficient, with about half of the initial ligands dissociated upon exposure. Our data reveal that the such dissociation can produce different outcomes, depending on the structural configuration. While all the organic groups were expected to detach and desorb from the resist in their entirety, a sizeable amount of them remain and form undesired byproducts such as alkene chains. In the framework of the materials synthesis and engineering through specific building blocks, these results provide a deeper insight into the photochemistry of resists, in particular for extreme ultraviolet lithography.

**Introduction**

Extreme ultraviolet (EUV) lithography is set to become the next generation lithography process for high volume manufacturing of semiconductor devices.[1] It represents a significant advancement over the current technology based on 193-nm wavelength because it uses light at the EUV wavelength of 13.5 nm. To realize EUV lithography, both industry and academia are undertaking significant effort in the development of exposure tools, optics, masks and resists.[2] The challenges and aims of resist development for EUV is mostly focused on the assessment of the patterning performance in terms of sensitivity, roughness, and resolution.[3]

Hybrid organic-inorganic resists are a recently developed class of lithographic materials which have a relevant technological importance for use as directly-patternable hard masks for pattern transfer applications.[4] In some of these materials, exposure to photon radiation, electron beam or ion beam, triggers the detachment of the organic ligand(s) while the inorganic core condenses to form a glassy network. The resulting oxide is insoluble to the developer, while the unexposed parts are removed in a process which effectively realizes a negative tone patterning. The combination of high resolution, high etch resistance, and compatibility with existing processes makes hybrid resists of great technological interest not only for small scale applications, but especially for large scale manufacturing using EUV lithography. For these reasons, extensive research is underway to synthesize photosensitive condensed metal oxide resists for EUV based on metal-oxide organic clusters,[5] metal-oxide cores,[6] and metallic oxoclusters.[7] Because absorption at EUV is entirely dictated by elemental composition, incorporation of transition metals and metals or semimetals with large atomic number whose *d* orbitals have large interaction cross section with EUV photons[8] brings the additional advantage of enhancing absorption and improving lithographic sensitivity.[9]

A comprehensive understanding of the physicochemical reactions that take place in a resist during lithographic patterning by exposure to EUV photons (92 eV) is still lacking. In particular, tracking the chemical changes in a resist during exposure is considerably challenging[10] even in the simplest molecules. Typical characterization methods include scanning electron microscopy, X-ray diffraction, X-ray photoelectron spectroscopy, secondary ion mass spectrometry and infrared spectroscopy.[2,11,12] The main disadvantages of these techniques lie in the need for large area exposure and in the time delay between exposure and analysis. Moreover, because EUV resists are usually spin coated in thin films (< 100 nm), the effect of adsorption, desorption, and contamination from ambient environment has been demonstrated to be not negligible.[7,11]

In this work, we analyzed the chemical changes occurring in hybrid organic-inorganic resists synthesized from organically modified precursors and transition metal alkoxides via the sol-gel route. These systems are designed to be patterned by ultraviolet, EUV and electron beam lithography and are based on oxides of Si, Ti and Zr, which have high mechanical resistance to anisotropic etch for pattern transfer applications. The analysis was carried out by near edge X-ray absorption fine structure (NEXAFS) spectroscopy, a technique that exploits the photoexcitation of electrons from a core level to unoccupied molecular orbitals to probe the specific chemical environment of a given element. Among other applications, this technique is appropriate to study the chemical properties in photodegradable materials, specifically those containing organic components. For instance, NEXAFS clarified the dissociation mechanism upon exposure of poly(methyl methacrylate) both qualitatively and quantitatively, disclosing which carbon bonds are sensitive to exposure and to what extent.[13] NEXAFS at the carbon edge is particularly suitable to analyze the chemical environment of the organic ligands of hybrid resists, regardless of the inorganic component.[14] This work thus aims to quantitatively assess the chemical changes in hybrid resists after exposure using the variation of optical density of specific resonances as signatures. Notably, the possibility of performing *in situ* exposure of the material is a key advantage over other techniques.

**Experimental**

Hybrid organic-inorganic resists were synthesized by sol-gel chemistry: hydrolysis and condensation reactions occur at temperature $T < 200$ °C ('chimie douce') according to methods described elsewhere.[15,16,17] These materials are nonchemically amplified synthesized from organically modified precursors in which organic groups can behave as network formers and/or network modifiers. The molecular formula of these modified alkoxides is $R_nSi(OR')_{4-n}$, with n = 1, 2. The organic network modifier functionalities consist of a non-hydrolizable organic group (alkyle, methyl, phenyl), while the organic network former contains polymerizable groups able to form an organic network covalently linked to the inorganic one. In this case, the inorganic network consisted of silicon, titanium and/or zirconium four-fold coordinated with oxygen. As a result, the resist after synthesis consists of an inorganic glassy network with covalently linked organic moieties, schematically shown in Fig. 1.

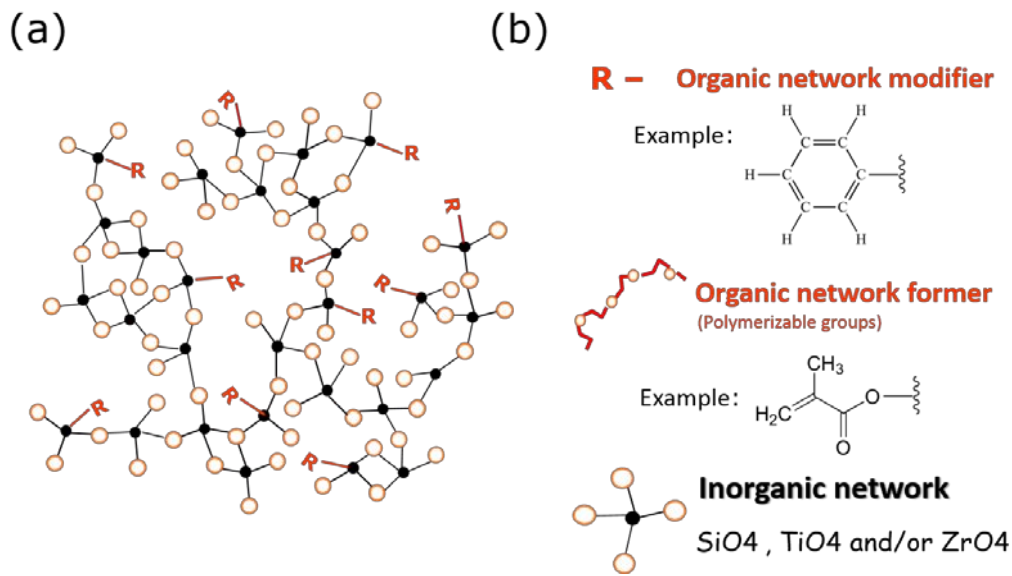

Figure 1. (a) Schematic representation of the chemical structure of the hybrid organic-inorganic resists after synthesis and before exposure. Symbols represent: silicon, zirconium or titanium (black circles), oxygen (orange open circles), organic groups (red '**R**'), organic ligands (red solid lines). (b) The components of the hybrid resist are: organic network modifier, where **R** is a non hydrolizable group (alkyle, methylalkyl, phenyl), organic network former where **R** is a polymerizable group, and inorganic network unit consisting of silicon, titanium and/or zirconium (solid black circle) coordinated by oxygen atoms (orange open circle).

Three varieties of hybrid resists have been investigated in this work by NEXAFS: their constituents are summarized in Table I. Sample S1 is a silicon oxide-based material with a phenyl bridging groups connecting two Si atoms; sample S2 has silicon oxide and titanium oxide as inorganic components, and phenyl pendant functionalities; sample S3 consists of an inorganic network of silicon oxide and zirconium oxide with methyl methacrylate as pendant organic groups. All the materials were spin-coated on thin (≈ 100 nm) silicon nitride membranes suspended on a silicon frame. After coating, a bake at 80 °C for 5 minutes was carried out to remove the casting solvent from the film. The resulting resist thickness was ≈ 100 nm, which is sufficient to achieve a measurable signal-to-noise ratio in the absorption spectra in transmission mode. The lithographic performance of these materials was assessed by electron beam lithography in previous works.[15,16,17]

Table I. Description of the constituents of the three hybrid organic-inorganic resists studied in this work.

| Sample | Inorganic network | Organic group |
|---|---|---|
| **S1** | $SiO_X$ | phenyl bridge |
| **S2** | $SiO_X + TiO_X$ | phenyl pendant |
| **S3** | $SiO_X + ZrO_X$ | methyl methacrylate |

NEXAFS experiments were conducted at the PolLux beamline[18,19] of the Swiss Light Source in the proximity of the carbon edge with an energy range of 280 – 350 eV and a calibrated accuracy better than 1 eV. The PolLux beamline features an STXM (scanning transmission X-ray microscope) which combines spatially resolved microscopy and NEXAFS spectroscopy.[20] In the STXM, the X-rays are focused by a Fresnel zone plate which enables the acquisition of absorption maps spatially resolved to about 20 nm. Although the spatial morphology is not of interest in the scope of this work (because spin coated resist films are chemically isotropic), the STXM makes it possible to locally expose a given area of the resist and to perform spectroscopy directly afterwards without further handling of the sample.

First, NEXAFS spectra of the pristine samples were acquired while scanning the defocused beam over a 1 μm long strip of resist, so as to prevent undesired exposure of the material itself. Afterwards, selective exposure of the material was done on an approximately round shape area of about 1 μm diameter. For this purpose, as EUV light of 13.5 nm wavelength is not covered by the energy range of Pollux, the photon energy was set to 500 eV. This energy was chosen because it does not match any known resonance of the material.[21] As a result, the exposure is expected to proceed by the generation of primary and secondary photoelectrons, in a similar fashion to what happens in lithographic processes carried out at an energy above the ionization potential of the resist (i.e., EUV, electron beam, ion beam). Being aware that irradiating with EUV light at 91.9 eV or soft X-rays at 500 eV might come along with some differences, the exposure *in situ* brings the remarkable advantage of keeping the sample in vacuum, preventing contamination from ambient (absorption, desorption) and reducing the delay between exposure and analysis which have been demonstrated to alter irreversibly the chemistry of the material.[7,11] Because the resist material absorbs weakly at 500 eV and owing to the relatively low flux of the PolLux in comparison to dedicated lithographic tools, the exposure time of even such a limited area was considerably long (10 min) by lithographic standards.

The identification of the position and intensity of the features of the absorption spectra was carried out by fitting a combination of mathematical functions to account for the physical mechanisms of soft X-ray absorption in the resist. Based on established literature of NEXAFS studies of organic compounds in the proximity of the carbon edge, the following phenomena were considered: I) photoexcitation of a K-shell electron to unoccupied $\pi^*$ or $\sigma^*$ molecular orbitals; II) photoexcitation of a K-shell electron to continuum states above the ionization potential (IP); and III) decreasing absorption in the extended structure (energy $E >$ IP). The fitting function $f(E)$ of the photon energy thus consisted of a combination of: 1) sum of up to 9 symmetric Gaussian functions; 2) one normal cumulative function centered at an energy equal to the ionization potential; and 3) an exponentially decaying term for the extended region:

$$f(E) = \sum_{i=1}^{9} \frac{A_i}{\sigma_i \sqrt{2\pi}} e^{-\frac{(E-\mu_i)^2}{2\sigma_i^2}} + A e^{-\frac{E-IP}{\sigma}} \cdot \frac{1}{\sigma_c \sqrt{2\pi}} e^{-\frac{(E-IP)^2}{2\sigma_c^2}}$$

(Eq. 1)

where the $A_i$ and $\sigma_i$ terms account for the heights and widths of the Gaussians, the $\mu_i$ terms account for the position of the Gaussian peaks, $\sigma_c$ accounts for the continuum states, and $A$ and $\sigma$ account for the magnitude and slope of the exponentially decaying term, respectively. The nonlinear least-squares fitting of $f(E)$ to the experimental data was carried out by the Levenberg-Marquardt algorithm. The goodness of the fit was guaranteed by iterating the algorithm until a coefficient of determination $R^2$ of 0.975 or better was achieved for the fitting of each experimental spectrum. Initial guesses for the location of the peaks of the Gaussian components were put in by visual inspection of the data. Finally, the ionization potential of the exposed samples was set equal to the IP value previously extracted in the unexposed condition.

## Results and discussion

The measured optical density (O.D.) of the resist films and the fitting functions to the data, before and after exposure, are shown in Fig. 2. In general, the attribution of resonances (peaks) to clearly identifiable electronic transitions across the entire NEXAFS spectrum of complex macromolecules is not straightforward. For the identification and interpretation of the peaks, in the following we refer to previous NEXAFS studies of organic compounds at the carbon edge.[13,22,23,24,25,26] To evaluate the relative change in the carbon environment at specific resonances, the height of the spectra in Fig. 2 was normalized to the respective O.D. at 350 eV. The value of O.D. at 350 eV is also reported in the following, to the purpose of estimating the total amount of carbon in the material before and after exposure.

The NEXAFS spectra (Fig. 2) of all samples show several common features. Primarily, the sharp peak at ~285 eV is clearly belonging to the $\pi^*$ C=C resonance of the carbon-carbon double bonds, present in both the phenyl groups (S1 and S2) and the methyl methacrylate (S3). Based on the chemical composition of these materials, the many overlapping peaks at 287 – 288 eV can be associated with resonances arising from carbon-hydrogen bonds (S1 and S2) typical of unsaturated phenyl rings. The sharp peak at 288 eV of sample S3 is attributed to C=O double bonds in the methyl methacrylate group. The large increase in the optical density above about 288 eV is, for all samples, a consequence of electronic transitions from carbon K-shell into states above the ionization potential (IP). Several additional peaks, beyond the IP, are indicative of σ* resonances owing to the presence of various C–C bonds, although their exact determination is not straightforward and goes beyond the scope of this work.

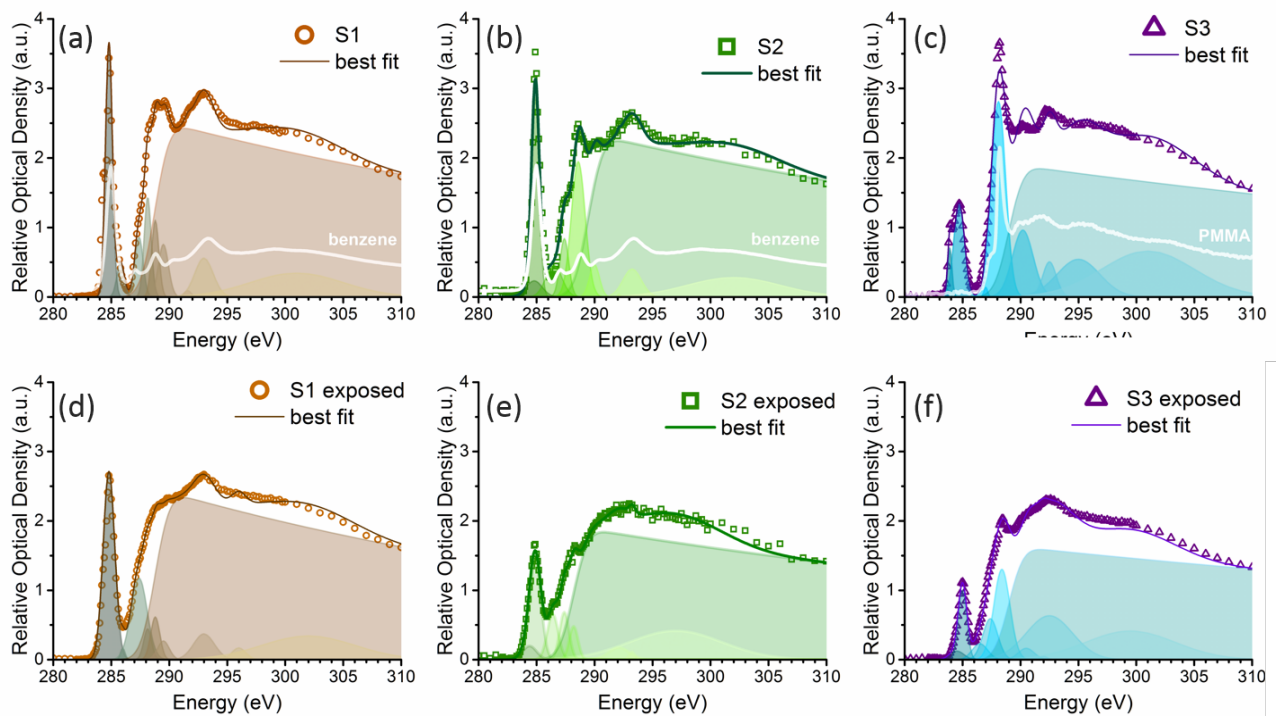

Figure 2. Normalized NEXAFS spectra (symbols) of hybrid organic-inorganic resist before exposure (a, b, c) and after exposure (d, e, f) of samples S1, S2, and S3, respectively. Also shown are the best fit (solid lines) and their individual components (shaded areas) obtained using the fitting function described in Eq. (1). For interpretation purposes, the NEXAFS spectra of benzene (a, b) and poly methyl methacrylate (c) are also shown (white line).

It can be observed that the spectra of both pristine samples S1 and S2 exhibit strong similarities to the spectrum of benzene[27] (also shown in Figs. 3a and 3b by the white line) owing to the presence of phenyl groups in the organic component. The difference in relative magnitude of the benzene peaks at 285, 289 and 293.5 eV in the reference spectrum is also a consequence of the anisotropy in the X-ray absorption of oriented benzene molecules (linear dichroism); while the NEXAFS spectra of our resist materials show no anisotropy due to the fully disordered structure of the spin coated sol-gel films. After exposure, the characteristic peaks of benzene are still present in S1 and S2 with weaker intensity than they were in the pristine material, indicating a modification of the organic component.

The NEXAFS spectrum of sample S3 shows several characteristic features in common with that of poly methyl methacrylate (PMMA), again indicated by the white line in Fig. 3c.[27] This similarity can be understood because of the identical nearest neighbor environment of one carbon atom in the ligand of S3 and in PMMA. It should be noted that the additional π* resonance at ≈ 285 eV in sample S3 is indicative of the carbon-carbon double bond which does not belong to the polymeric form (PMMA) but it is characteristic of the methyl methacrylate monomer. After exposure, the intensity of the 288 eV peak of the π* C=O double bond decreases significantly, which is also ascribed to the dissociation of the methyl methacrylate group.

To clarify how the exposure of a hybrid organic-inorganic resist proceeds, i.e. what changes occur in the carbon environment, we associate the variation of optical density of specific resonances of the NEXAFS spectra to changes in the phenyl bridged (S1), phenyl pendant (S2) or methyl methacrylate (S3) organic groups. We consider both the π* C=C resonance (285 eV) and one of the benzene resonance (293.5 eV) as markers for samples S1 and S2, while the π* C=O resonance (288.5 eV) is taken as the most relevant in the case of sample S3. Because the organic group can dissociate in a variety of ways upon exposure, the total amount of organic component needs to be accounted for. To this purpose, the unnormalized optical density at 350 eV (i.e. at energy well beyond any other resonance of the carbon edge) is taken as a marker for the total carbon content of the film. The experimental values of all these indicators are listed in Table II.

After exposure, the signature of benzene in sample S1 (bridging phenyl) was preserved more markedly than it was in sample S2 (pendant phenyl). Quantitatively (Table II), this observation was confirmed by the small decrease of intensity of the 293.5 eV resonance in the former (- 25 %) in comparison to the latter (- 67 %). The loss of carbon-carbon double bonds, indicated by the change in the intensity of the 285 eV resonance, followed a similar trend to that of carbon atoms in the benzene ring configuration. It can be concluded that the phenyl group is more easily dissociated when it is in the pendant configuration than when in the bridging configuration. However, in absolute terms, sample S1 lost more organic component (- 38 %) than sample S2 did (- 32 %) after exposure. The outcome of the decomposition of the phenyl ring was, therefore, different in these two cases. In sample S1, the total amount of carbon lost was greater than the relative amount of dissociated phenyl groups, indicating that the loss of carbon upon exposure was also caused by the detachment and desorption of the entire organic group from the film. On the other hand, while the phenyl decomposition occurred more efficiently in sample S2, it resulted in only a limited desorption from the film. In this case, the retention of a significant amount of carbon is ascribed to the formation of organic byproducts such as the formation of alkenes and/or polyenes.

In the case of the methyl methacrylate ligand of sample S3, a sizeable degradation of the methacrylate group occurred in more than half of the double carbon-oxygen bonds initially present in the material. However, the remaining strongly bonded carbon did not decrease as much as expected (- 15 %) and only one third of the initial carbon atoms were lost after exposure. As in the case of the phenyl pendant (S2), this observation suggests that not all of the dissociated methacrylate groups were desorbed. Based on the structure of the resist, it is thought that the release of oxygen occurs by dissociation of the carbon-oxygen bond,[26] while the remaining carbon bonds in all other configurations (such as, bridging two metal atoms or one metal and one oxygen atom).

Table II. Normalized optical density (O.D.) of hybrid organic-inorganic resists at characteristic NEXAFS resonances of 285, 293 and 288.5 eV in the initial condition, after exposure, and relative variation. The O.D. at 350 eV is the unnormalized height, indicative of total carbon content in the material.

| Energy [eV] | Optical Density | | | | | | | | |
|---|---|---|---|---|---|---|---|---|---|
| | S1 | | | S2 | | | S3 | | |
| | initial | exposed | Δ | initial | exposed | Δ | initial | exposed | Δ |
| 285 π* C=C | 3.57 | 2.72 | **- 24 %** | 3.09 | 1.52 | **- 51 %** | 1.31 | 1.10 | **- 15 %** |
| 293.5 phenyl | 0.49 | 0.37 | **- 25 %** | 0.31 | 0.10 | **- 67 %** | – | – | – |
| 288.5 π* C=O | – | – | – | – | – | – | 2.77 | 1.33 | **- 52 %** |
| 350 (unnormalized) total carbon | 0.60 | 0.37 | **- 38 %** | 0.64 | 0.43 | **- 32 %** | 0.18 | 0.12 | **- 33 %** |

The efficiency of dissociation of hybrid photoresists, here measured by NEXAFS, provides useful insight into the lithographic energy efficiency of these materials. The bridged phenyl configuration resulted in a lower sensitivity to exposure: therefore, a higher dose would be required for patterning (also confirmed by independent exposures performed by electron beam lithography).[16] On the other hand, a lower sensitivity brings the benefit of higher stability and longer shelf life of the material which, after synthesis, is less prone to agglomeration. Besides, the finding that a considerable amount of doubly bonded carbon was still preserved or even newly formed after exposure is highly relevant for lithographic purposes. While the ideal outcome of the lithographic exposure of a hybrid resist is typically considered to be an amorphous metal oxide, the presence of carbon after exposure is not detrimental in itself, as the exposed resist is meant to be used as a hard mask for pattern transfer. For instance, fullerene-based resists for electron beam and EUV lithography exploit the network of single- and double-bonded carbon for its high mechanical stability and etch resistance.[28] There are other cases, such as in selective electron-beam induced deposition of Pt from a gaseous organometallic precursor, where carbon contamination has a detrimental effect on the mechanical and electrical properties of the pattern.[29] Previous studies ascribed the formation of carbonaceous residue to the metalorganic precursors in the gas phase when patterning and fabrication of nanostructures was carried by electron-beam[29] and soft X-rays alike.[30] Further research is required to conclusively determine the effect of carbon residue for lithographic applications and pattern transfer by deep anisotropic etch.

## Conclusions

In summary, NEXAFS is a viable a tool for the study of chemical changes in hybrid organic-inorganic resists. Based on established works from literature, we verified the chemical composition of the organic part as containing specific groups (either phenyl or methyl methacrylate). Exploiting the high chemical selectivity of this technique, we tracked the changes in the organic component upon exposure, regardless of the contribution of the inorganic glassy network. It was found that exposure to soft X-rays is effective in dissociating the organic groups similarly to what happens in EUV and electron beam lithography. However, the outcome of the dissociation is not always the desired complete detachment and desorption of the organic group. We used the relative intensity of the doubly bonded carbon resonance, of the doubly bonded carbon-oxygen resonance and of the benzene ring as signatures of the chemical changes in the organic part of the material. Interestingly, the dissociation efficiency of the organic ligand was found to be dependent on its chemical configuration and several different dissociation pathways are possible. The persistence of strongly bonded C=C in the exposed material is a finding of interest to lithographers who wish to achieve pattern transfer by physical etching. These results provide insight into the mechanism by which hybrid organic-inorganic resists work, which can open new opportunities for their chemical synthesis for performance improvement.

## Conflicts of interest

There are no conflicts to declare.

## ACKNOWLEDGEMENTS

Part of this work was conducted at the Swiss Light Source. The PolLux end station was financed by the German Minister für Bildung und Forschung (BMBF), contract 05 KS4WE1/6. This study has received funding from the EU-H2020 Research and Innovation program under Grant Agreement No. 654360 NFFA-Europe.